\documentclass[12pt]{article}

\usepackage{color,soul}
\usepackage{graphicx}
\usepackage[font=small,labelfont=bf]{caption}

\usepackage{amsmath,amssymb,amsthm}

\usepackage{hyperref}

\begin{document}
	
	\begin{center}
		
		{\bf \large On possible tachyonic state of neutrino dark matter}
		\vskip20pt{M. A. Makukov$^1$, E. G. Mychelkin$^1$ and V. L. Saveliev$^2$
			\vskip9pt
			{\footnotesize
				{\it $^1$Fesenkov Astrophysical Institute,\\
				 $^2$Institute of Ionosphere,\\
				 National Center of Space Research and Technology,\\
				 Almaty 050020, Republic of Kazakhstan\\}}
				\vskip9pt
				\footnotesize{E-mail: makukov@aphi.kz (M.A.M.), mychelkin@aphi.kz (E.G.M.)}
		}
	\end{center}
	
%
%
\vskip16pt

\begin{abstract}
 We revive the historically first dark matter model based on neutrinos, but with an additional assumption that neutrinos might exist in tachyonic almost sterile states. To this end, we propose a group-theoretical algorithm for the description of tachyons. The key point is that we employ a distinct tachyon Lorentz group with new (superluminal) parametrization which does not lead to violation of causality and unitarity. Our dark matter model represents effectively scalar tachyonic neutrino-antineutrino conglomerate.  Distributed all over the universe, such fluid behaves as stable isothermal/stiff medium which produces somewhat denser regions (`halos') around galaxies and clusters. To avoid the central singularity inherent to the isothermal profile, we apply a special smoothing algorithm which yields density distributions and rotation curves consistent with observational data.\\

 
\end{abstract}

\section{Introduction}
The possibility of tachyonic nature of neutrinos has been discussed for three decades. In experiments on parity violation in weak interactions neutrinos are always left (and antineutrinos are right). If neutrino velocities were less than the speed of light, in certain frames the neutrino helicity (the projection of the spin onto the direction of momentum) would inevitably swap to the opposite. This has never been observed, so the possible conclusion is that neutrino velocity must be equal or greater than the speed of light. But effects of neutrino oscillations show that neutrinos have small but non-zero masses and thus cannot travel at the speed of light. So, tachyonic nature of neutrinos might be considered as the consequence of chiral invariance rather than an \emph{ad hoc} hypothesis. However, the excess over the speed of light is expected to be so insignificant that it hardly can be measured directly in modern Earth-based experiments, and the implementation of the above-mentioned frames in experiment is practically impossible, so the problem remains to be open.\\
\indent In our approach, apart from the neutrinos originating from annihilation reactions of leptons at the temperatures of a few MeV during early stages of the universe evolution (`secondary neutrinos'), there exists a background of primordial tachyonic neutrinos which comprise the bulk of dark matter (DM). The expected contribution of all secondary neutrinos to DM, as well known, is only about 0.1\%.

\section{New Approach to Tachyons}
The customary approach to tachyons within special relativity as to spacelike objects with imaginary masses leads to violation of causality and/or unitarity, instability of tachyonic modes, negative energies, violation of the Pauli principle \cite{Feinberg1967}, lack of a generally accepted scheme of (second) quantization for tachyons, difficulties with the classification of tachyons by irreducible representations of the Poincar\'e group (impossibility to prescribe definite spin values to tachyons \cite{Bekaert2006}), etc.\\
\indent Nevertheless, there is a growing tendency of various data in favour of tachyonity of neutrinos \cite{Ehrlich2015}. Besides, it is relevant to note that neutrino signal ($E(\nu_e)\approx 10$ MeV) from supernova SN1987A was registered a few hours before the optical one, suggesting that  neutrino velocity excess is $\Delta c/c \le 10^{-9}$ at such energies. 

\indent To overcome the aforementioned obstacles we employ the squared velocity inversion algorithm applied to squared components of 4-momenta in the dispersion relation (with $c=1$): $v^2=1/u^2 \rightleftarrows u^2=1/v^2, 0\le v<1, 1<u<\infty$. Then
\begin{equation}
	E^{2} (v)-p^{2} (v)=m^{2} \quad \rightleftarrows \quad \tilde{p}^{2} (u)-\tilde{E}^{2}(u)=m^{2},
	\label{dispersion}
\end{equation}
where bradyon and tachyon energies are $E^2(v)=m^2/(1-v^2)$ and  $\tilde{E}^2(u)=m^2/({{u}^{2}}-1),$
and similarly for momenta.\\
\indent This reparametrization algorithm implies that for both relations in (\ref{dispersion}) the sign of the mass-term remains positive and thus does not lead to negative energies and violation of causality/unitarity, but the signature of the Minkowski-type space for tachyons changes to the opposite. Since the metric is an invariant of the Lorentz group, the change of the signature implies transition to another group.\\ 
\indent Thus we now work within a new `tachyon Lorentz group' which has another (superluminal) parametrization and acts on Minkowski space possessing the opposite signature. The group parameters $v=|\mathbf{v}|$ and $u=|\mathbf{u}|$ connected via aforementioned non-isomorphic mapping are, strictly speaking, related to different 4-spaces with different Euclidean subspaces. Similar inversion algorithm was proposed by Recami \cite{Recami1977} but within a single Euclidean subspace which is still somewhat problematic.

\indent Considering such extension of the Lorentz group for helicity-conserving tachyons, we must also replace Dirac-conjugated wave functions $\bar{\psi }=\psi ^{\dag } \gamma _{0} $ with Hermitian-conjugated ones $\psi ^{\dag } $ in tachyon neutrino sector. Then the neutrino mass-term will be well-defined separately for neutrinos and antineutrinos (this is impossible in case of the standard Dirac conjugation), and a new tachyon Dirac equation arises: $\left(i\gamma ^{\mu } \partial _{\mu } -\Gamma m\right)\psi =0$, $\Gamma =\gamma _{0} \gamma _{5} $, $\Gamma ^{2} =-1$, which can be split into two independent equations,  $\left(p_{0} +\vec{\sigma }\vec{p}+m\right)\psi _{R} =0$ and $\left(p_{0} -\vec{\sigma }\vec{p}-m\right)\psi _{L} =0$, separately for ($R$-right) antineutrinos and ($L$-left) neutrinos. As a result, superposition of the squares of free tachyon spinor neutrino $\nu $ and antineutrino $\bar{\nu }$ fields represents a scalar conglomerate $\Phi =\psi ^{\dag } \psi =\nu ^{2} +\bar{\nu }^{2} $, $\psi =\nu +i\bar{\nu }$. Such effectively scalar field in the Einstein and Klein-Gordon equations leads to reasonable explanation of the dark matter (DM) phenomenon as the primordial tachyon neutrino-antineutrino background \cite{Mych2009,MychMak2015}.

\section{Isothermal Dark Matter}
To obtain the approximate estimation for corresponding distribution of dark matter density we follow the traditional hydrodynamic approach, using the energy density $\varepsilon$ of dark matter instead of the mass density, in accord with the conventional relativistic condition of equilibrium reduced in weak gravitational field \cite{Landau} to:
\begin{equation}
	\nabla p=\frac{dp}{d{{x}^{i}}}=-\frac{\varepsilon }{2}\frac{\text{d}}{\text{d}{{x}^{i}}}\ln {{g}_{00}}\approx -\frac{\varepsilon }{2}\frac{\text{d}}{\text{d}{{x}^{i}}}\left( 1+2\varphi /{{c}^{2}} \right),
	\label{HydroEquil}	
\end{equation}		            
where $p$ is pressure, ${{g}_{00}}$ is the metric tensor component. In spherically symmetric case for both tachyon and bradyon gas with approximately barotropic (isothermal) equation of state $p\approx w\varepsilon $ we get
\begin{equation}
	\frac{d}{dr}\left( \frac{{{r}^{2}}}{\varepsilon }\frac{d\varepsilon }{dr} \right)+\frac{4\pi G}{w{{c}^{4}}}{{r}^{2}}\varepsilon = 0.
	\label{BarotropDiffEq}				   
\end{equation}		            
Just as in non-relativistic case, this equation has a particular solution known as the singular isothermal sphere: $\varepsilon (r)=2/B{{r}^{2}}$  with $B\approx 4\pi G/w{{c}^{4}}$. By an order of magnitude the factor $B^{-1}$ might be interpreted as the ratio of mass-energy to corresponding length scale $\lambda: {{B}^{-1}}=M{{c}^{2}}/4\pi \lambda$.\\
\indent It might be shown that for usual (bradyonic) matter, the typical equations of state have $0 \leq w \leq 1/3$, while tachyonic matter, in principle, admits $-1 \leq w \leq 1$ (excluding the aforementioned bradyonic interval $0 \leq w \leq 1/3$).  As shown, e.g., in \cite{Lukacs}, even in strong fields the general relativistic stiff equation of state for self-gravitating systems can also induce isothermal behavior of matter.\\
\indent Despite inappropriate asymptotic behaviour, the isothermal (constant temperature) law seems to be natural specifically in tachyonic case due to extremely weak interaction of such DM with ordinary matter. For large $r$ such solution leads to saturated rotation curves and thus is observationally favored. Then for the Poisson equation for a point source on the isothermal DM background of some scale $\lambda $ related to effective halo mass as $G{{M}_{DM}}/\lambda =V_{0}^{2}$, where ${{V}_{0}}$ is the appropriate amplitude of the rotation curve, we get:
\begin{equation}
	{{\nabla }^{2}}\Phi =4\pi G{{\rho }_{tot}}(r)=4\pi G[{{\rho }_{N}}(r)+{{\rho }_{DM}}(r)]=4\pi G\left[ {{M}_{N}}\delta (r)+\frac{{{M}_{DM}}}{4\pi \lambda }\frac{1}{{{r}^{2}}} \right].
	\label{Poisson}
\end{equation}
It has a solution representing a sum of the singular Newtonian (‘baryon’) and DM-background terms with effect of saturation of rotation curves for sufficiently large scales:
\begin{equation}
	\Phi =-\frac{G{{M}_{N}}}{r}+\frac{G{{M}_{DM}}}{\lambda }\ln \left( \frac{r}{\lambda } \right).
	\label{Pot}
\end{equation}
Unlike commonly used phenomenological density profiles for dark matter halos (see, e.g., in \cite{Binney2008}), isothermal sphere profile has fundamental physical justification. However, it is still not fully realistic due to singular behavior at the center.
\section{Smoothing Algorithm}
To avoid problems with the singularities we apply the following smoothing algorithm, replacing point and singular distributions with smoothed ones. For that, we apply to both sides of the Poisson equation (\ref{Poisson}) the smoothing operator $\mathsf{\hat{S}}$ defined as a convolution of normalized Gaussian distribution with a function in question $f(\mathbf{r})$:
\begin{equation}
	\mathsf{\hat{S}}f(\textbf{r})=\hat{f}(\textbf{r})={{\left( 2\pi \sigma  \right)}^{-{\scriptstyle{}^{3}/{}_{2}}}}\int{{{e}^{-\frac{{{(\textbf{r}-\textbf{r}')}^{2}}}{2\sigma }}}}f(\textbf{r}'){{d}^{3}}\textbf{r}'.
	\label{SmoothOp}	
\end{equation}
In case of spherical symmetry this operator reduces to one-dimensional integral
\begin{equation}
	\mathsf{\hat{S}}f(r)=\hat{f}(r)={{\left( 2\pi \sigma  \right)}^{-{\scriptstyle{}^{1}/{}_{2}}}}{{r}^{-1}}\int\limits_{0}^{\infty }{\left[ {{e}^{-\frac{{{\left( r-r' \right)}^{2}}}{2\sigma }}}-{{e}^{-\frac{{{\left( r+r' \right)}^{2}}}{2\sigma }}} \right]f(r')r'dr'}.
	\label{SmoothOpSpher}
\end{equation}
In these expressions caps stand for averaging, $\hat{f}={{\left\langle f \right\rangle }_{\sigma }}$, and at the limit $\sigma \to 0$ we return to initial singular (unsmoothed) sources. Acting with this operator on the density distribution of a point source and singular isothermal sphere (right side of the equation (\ref{Poisson})), we get the smoothed point source and the smoothed isothermal sphere (see Fig.~\ref{f1}):
\begin{figure}[pb]
	\centerline{\includegraphics[width=12cm]{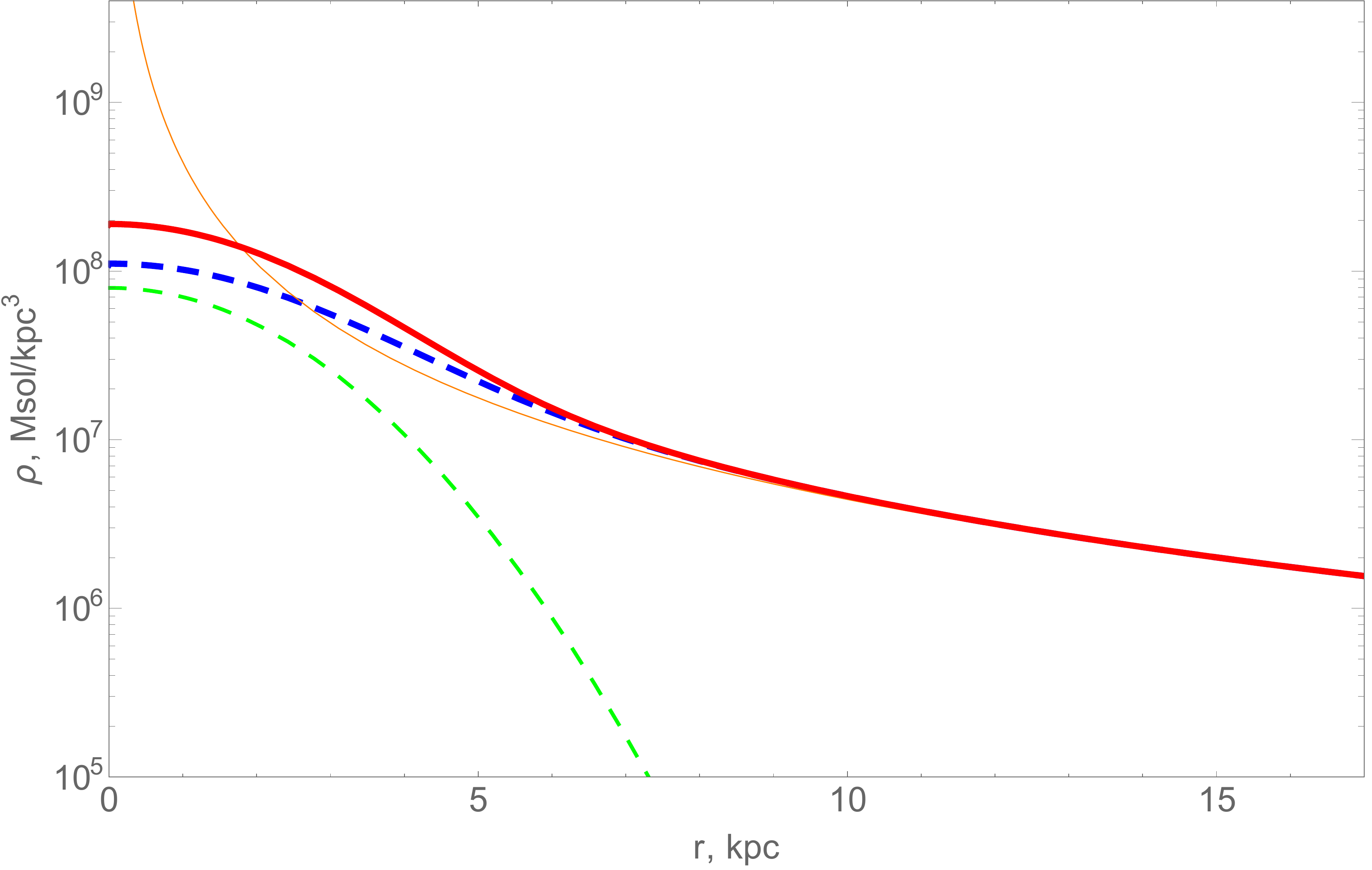}}
	\vspace*{8pt}
	\caption{Typical density distributions representing smoothed point source (green curve) and isothermal sphere (blue curve). The unsmoothed singular isothermal distribution (orange curve) is also shown for comparison. Red curve stands for the total smoothed density. The parameters are chosen to fit observational data in Fig.~\ref{f2}. \label{f1}}
\end{figure}
\begin{equation}
	{{\hat{\rho }}_{tot}}(r)={{\left( 2\pi \sigma  \right)}^{-3/2}}{{M}_{N}}{{e}^{-{{r}^{2}}/(2\sigma )}}+{{(4\pi \sigma \lambda )}^{-1}}\sqrt{2\sigma }{{M}_{DM}}{{r}^{-1}}F\left( r/\sqrt{2\sigma } \right),
	\label{SmoothedDensity}
\end{equation}
where $F(x)$ is the Dawson function (note that in spherical coordinates $\delta (r)\to \delta (r)/(2\pi {{r}^{2}})$). The potential corresponding to this smoothed density distribution might be obtained either by solving the Poisson equation for the smoothed distributions, or, taking into account the commutativity of Laplacian and convolution operators, acting directly with $\mathsf{\hat{S}}$ onto (\ref{Pot}). Both approaches yield identical results (up to a constant) which have the form
\begin{equation}
	\hat{\Phi }(r)=-G{{M}_{N}}\frac{\text{erf}\tfrac{r}{\sqrt{2\sigma }}}{r}+\frac{G{{M}_{DM}}}{6\lambda \sigma }{{r}^{2}}\times{}_{2}{{F}_{2}}\left( \left. \begin{matrix}
		1 & 1  \\
		2 & 5/2  \\
	\end{matrix} \right|;-{{r}^{2}}/2\sigma  \right),
	\label{SmoothedPot}
\end{equation}
where ${}_{2}{{F}_{2}}\left( \left. \begin{matrix}
a & b  \\
c & d  \\
\end{matrix} \right|;z \right)$ is the generalized hypergeometric function.\\

\begin{figure}[pb]
	\centerline{\includegraphics[width=12.0cm]{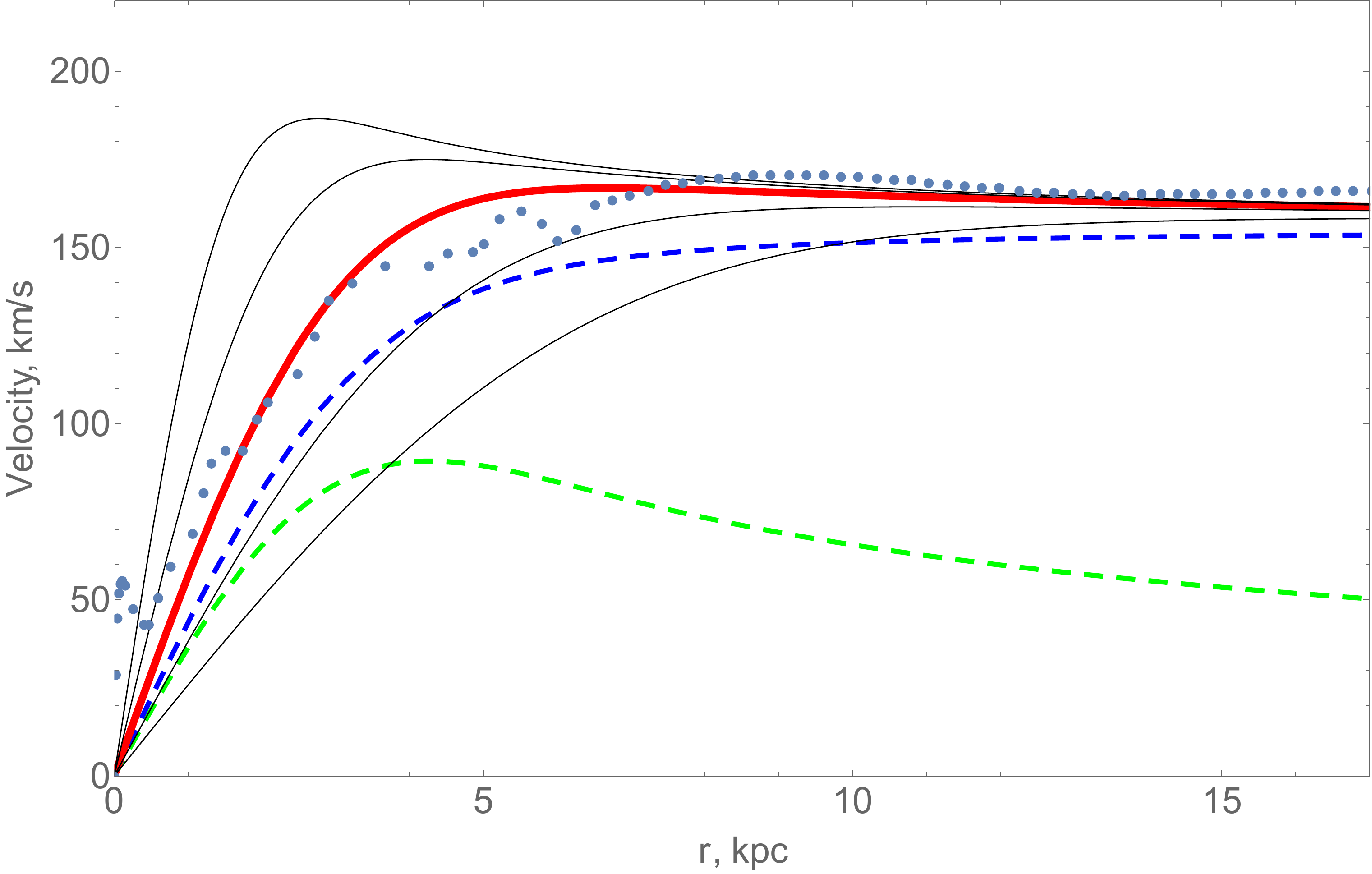}}
	\vspace*{8pt}
	\caption{Rotation curves for smoothed Newtonian (green), logarithmic (blue) and total (red) potentials. The dots represent observational rotation curve for the NGC3198 galaxy, and the  parameters for theoretical curves are chosen to fit it: ${{M}_{N}}={{10}^{10}}{{M}_{\odot }}$, ${{M}_{DM}}=20{{M}_{N}}$, $\sigma = 4$ kpc$^2$, $\lambda$ = 36 kpc. Thin black lines represent total rotation curves for different values of $\sigma$ (from the upper curve to the lower one): 1, 2, 8 and 16 kpc. \label{f2}}
\end{figure}
\indent The rotation curves corresponding to potential (\ref{SmoothedPot}) are:
\begin{equation}
	{{V}^{2}}(r)=-r\frac{d\hat{\Phi }}{dr}=G{{M}_{N}}\left( {{r}^{-1}}\text{erf}\frac{r}{\sqrt{2\sigma }}-\sqrt{\frac{2}{\pi \sigma }}{{e}^{-\frac{{{r}^{2}}}{2\sigma }}} \right)+\frac{{{M}_{DM}}G}{r\lambda }\left( r-\sqrt{2\sigma }F\left( r/\sqrt{2\sigma } \right) \right),
	\label{RotCurve}    
\end{equation}
This expression as a sum of baryonic (Newtonian) and DM (isothermal) terms is sufficiently flexible. The particular realization corresponding to densities in Fig.~\ref{f1} is shown in Fig.~\ref{f2} in comparison with observational data for NGC3198 \cite{Sofue}. Thus, the applied smoothing algorithm resulting in the second term in (\ref{SmoothedDensity}) proportional to ${{M}_{DM}}$ might be used to represent DM density distribution instead of commonly employed phenomenological halo profiles.
\section{Conclusions}
In our approach dark matter (DM) is represented by tachyonic neutrino-antineutrino conglomerate distributed all over the universe and accumulated around galaxies and clusters. At the current stage of the universe evolution tachyon neutrino DM background might be considered as a stationary fluid, which in our case is described on the basis of generalized isothermal halo profile leading to an appropriate additive term to the Newton-type potential. Employing the smoothing algorithm described above it is easy to obtain the observed saturated rotation curves by appropriate fit of smoothing parameters without invoking any {\it ad hoc} profiles.

As for the non-stationary initial hot stage of the universe evolution, the same gravitating neutrino-antineutrino conglomerate might, in principle, be regarded as a seed material for Pervushin's `dilaton fabric' producing intermediate vector bosons \cite{Pervushin}. This would be the case if colliding radial beams of tachyon neutrinos and antineutrinos in the central domain of super-strong gravitational field could be reprocessed into vector bosons and leptons, $\nu +\bar{\nu }\to W^{+} +W^{-} $, $\nu +\bar{\nu }\to Z$, $\nu +\bar{\nu }\to e^{+} +e^{-} $, with the subsequent universe evolution close to the standard scenario. 

It should be noted that existence of the primary tachyon neutrino DM background considered here does not imply by itself that the secondary neutrinos (produced at the cosmological temperatures about a few MeV from leptons annihilation) must be of tachyonic nature as well, and today we cannot exclude the possibility of production of the rest of neutrinos in bradyonic states. This might be established or disproved in future by comparing potential observations of cosmic neutrino background \cite{Delloro} with high- and low-energy Earth-based neutrino experiments and also by detecting neutrino flows from supernovae like SN-1987A within our Galaxy.

\end{document}